\documentclass[aps,prc,twocolumn,amsmath,showpacs,superscriptaddress]{revtex4}
\usepackage{graphicx,epsfig,longtable}
\usepackage{mathptmx}
\usepackage[light,first]{draftcopy} 
\usepackage{epstopdf}
\usepackage[pdftex]{hyperref}

\newcommand{\beqy}{\begin{eqnarray}}
\newcommand{\eeqy}{\end{eqnarray}}
\newcommand{\bmlet}{\begin{subequations}}
\newcommand{\emlet}{\end{subequations}}
\newcounter{saveeqn}

\def\gsimeq{\,\,\raise0.14em\hbox{$>$}\kern-0.76em\lower0.28em\hbox  
{$\sim$}\,\,}  
\def\lsimeq{\,\,\raise0.14em\hbox{$<$}\kern-0.76em\lower0.28em\hbox  
{$\sim$}\,\,}  

\begin{document}

\title{Impact of a low-energy enhancement in the $\gamma$-ray strength function on the radiative neutron-capture}

\author{A.~C.~Larsen}
\email{a.c.larsen@fys.uio.no}
\affiliation{Department of Physics, University of Oslo, N-0316 Oslo, Norway}
\author{S.~Goriely}
\affiliation{Institut d'Astronomie et d'Astrophysique, Universit\'e Libre de Bruxelles, CP 226,  1050 Brussels, Belgium}

\date{\today}

\begin{abstract}
A low-energy enhancement of the $\gamma$-ray strength function in several light and medium-mass nuclei has 
been observed recently in $^3$He-induced reactions. The effect of this enhancement on $(n,\gamma$) 
cross-sections is investigated for stable and unstable neutron-rich Fe, Mo and Cd isotopes. 
Our results indicate that the radiative neutron capture cross sections may increase considerably 
due to the low-energy enhancement when approaching the neutron drip line. This could have 
non-negligible consequences on r-process nucleosynthesis calculations.
\end{abstract}


\maketitle

\section{Introduction}

The $\gamma$-ray strength function or radiative strength function (RSF) characterizes 
average electromagnetic decay properties of excited nuclei. This quantity is important for describing 
the $\gamma$-emission channel in nuclear reactions. It is also indispensable for calculating nuclear 
reaction cross sections and reaction rates relevant for astrophysical applications. 
Recent studies~\cite{ar07,go98} clearly show the importance of a precise description of the 
$\gamma$-ray strength function at low energies, especially for a proper understanding of the 
nucleosynthesis of the elements heavier than iron by the rapid neutron-capture process (r-process). The r-process 
nucleosynthesis is called for to explain the origin of about half of the stable nuclides heavier 
than iron observed in nature and is believed to result from an extremely large neutron irradiation 
on timescales of the order of about one second. 

So far, however, the astrophysical site hosting such an r-process remains unknown. 
Although an $(n,\gamma$)--($\gamma,n$) equilibrium might take place in an environment with  
high neutron density and high temperature (in which case the r-abundance distribution remains 
rather insensitive to the reaction rates), more general r-process simulations require a reliable 
determination of the radiative neutron capture rates for all nuclei involved \cite{ar07}. Indeed, 
it should be kept in mind that even if the thermodynamic conditions of the r-process site remain unknown, 
the assumption of an $(n,\gamma$)--($\gamma,n$) equilibrium can only be tested if the neutron capture, 
$\beta$-decay and photodisintegration rates for all neutron-rich nuclei synthesized during such a process 
can be estimated reliably \cite{go96}. Furthermore, in specific sites such as the decompression of neutron 
star matter or the so-called cold neutrino-driven wind, neutron captures are in competition with 
$\beta$-decays, not with photodisintegrations, so that no $(n,\gamma$)--($\gamma,n$) equilibrium can 
be achieved and the final r-abundance distribution may well be sensitive to reaction rates. Finally, 
in any site, the final abundance distribution is likely to depend more or less  on the freeze-out 
conditions for which the $(n,\gamma$)--($\gamma,n$) competition comes out of equilibrium. 
Under such conditions, an accurate and reliable determination of all the ingredients of 
relevance in the calculation of the neutron capture rates, including in particular 
the $\gamma$-strength function, is required.

The nuclear physics group in Oslo has performed measurements on the $\gamma$-ray strength functions 
below neutron threshold of various light and medium-mass nuclei \cite{Fe_Alex,Fe_Emel,Mo_RSF,V,Sc}. 
These data have revealed an unexpected increase in the $\gamma$-decay probability at low $\gamma$-ray energies.  This enhancement is seen to be present typically for $E_\gamma \leq 3 $ MeV. In contrast to other soft resonances  observed previously such as the $M1$ scissors mode \cite{PragueM1,SchillerM1}, the physical origin of the  enhancement remains unknown. There is, for the time being, no established theory that is able to explain this \textit{upbend} phenomenon.

For nuclei close to the valley of stability, one might expect that the low-energy enhancement would have 
little influence on the neutron-capture cross section. Naturally, the most important energy region  
for the neutron-capture reaction is in the vicinity of the neutron separation energy $S_n$,
and low-lying structures in the $\gamma$-ray strength function would probably have a relatively small effect. 
However, for neutron-rich nuclei approaching the neutron drip line, the neutron separation energy rapidly decreases and enters the energy region where the enhanced strength is observed. 
This work aims at investigating how such a very low-energy strength enhancement may influence the 
neutron-capture cross section of exotic neutron-rich nuclei. Since this upbend phenomenon has been clearly seen in 
Mo isotopes, the present study will focus on Mo. The Mo case is also of particular interest due to the many observational 
constraints that can help us to determine the full RSF; these include the resonance spacing and the average total radiative 
width at the neutron separation energies, photoneutron cross-section data, and the measurements by the Oslo group.  

In Sect.~\ref{sect_exp}, we will describe the experimental results obtained by the Oslo method and the parameterizations 
used to model the low-energy $E1$ strength, in particular in the vicinity of the upbend structure. 
In Sect.~~\ref{sect_res}, the neutron-capture cross sections and the corresponding astrophysical rates are estimated 
for the Mo nuclei, as well as Fe and Cd isotopic chains. Finally, implications of these results and conclusions are 
discussed in Sect.~\ref{sect_conc}.

\section{Experimental and theoretical description of the upbend structure}
\label{sect_exp}

\subsection{Experimental results}
During the last decade, the Oslo group has developed an experimental method capable of extracting information 
on the nuclear level densities (NLD) and $\gamma$-ray strength functions by analyzing particle-$\gamma$ 
coincidence data from neutron pickup ($^3$He,$\alpha\gamma$) and inelastic scattering ($^3$He,$^3$He$^{\prime}$$\gamma$) 
reactions. Details about the method can be found in \cite{Schiller00}. 

One major result obtained through the Oslo method concerns an increase of the RSF at decreasing photon energy. 
This upbend structure has been observed in $^{44,45}$Sc, $^{50,51}$V, $^{56,57}$Fe, and $^{93-98}$Mo at energies 
typically smaller than 3~MeV. Heavier isotopes of Sn, Sm, Dy, Er and Yb for which similar experiments have been 
conducted do not show such a low-energy behavior. At the moment, only the Fe results have been 
confirmed by another experimental technique (the two-step cascade method, see~\cite{Fe_Alex}), and theoretically 
there is no model that can provide possible explanations. Even the multipolarity of the strength remains unknown. 

Guttormsen et al. \cite{Mo_RSF} showed that in the case of an $E1$ strength, the reduced strength in the $1-3$~MeV 
range would correspond to an average $B(E1)$ value of 0.02 $e^2$fm$^2$ (i.e. about 0.07\% of the $E1$ sum rule), 
in case of an $M1$, $B(M1)\simeq 2~\mu_N^2$ which is 3 to 4 times larger than the observed strength to 
mixed-symmetry $1^+$ states around 3~MeV~\cite{Fransen1,Fransen2}, and in case of an $E2$ strength, 
$B(E2)\simeq 15000~e^2$fm$^4$ which is 5 to 15 times the strength of the (de)excitation of the first $2^+$ 
states in the even Mo isotopes. 

Recent results on $^{60}$Ni investigated with the two-step cascade method applied on 
$(p,2\gamma)$ data~\cite{Alex_Ni}, indicate that there is an upbend in the $M1$ component
of the RSF and possibly also in the $E1$ component. However, one should note that this specific nucleus 
has only positive-parity states below $E_x \approx 4$ MeV, and one of the conclusions in Ref.~\cite{Alex_Ni} is
in fact that the low-energy enhancement is probably due to secondary $M1$ transitions in this excitation-energy region, while
the Oslo results reveal the RSF in the quasi-continuum region above $E_x \approx 4$ MeV.
Thus, the multipolarity and the electromagnetic character of the upbend in other cases are still unknown.
Although we cannot exclude any of the above-mentioned posibilities, we will  
assume that it can be associated with $E1$ $\gamma$-ray transitions for the nuclei studied in this work.

For the Mo case, additional information exists for the strength function below the neutron threshold. It concerns the 
measured $E1$ strength for $^{93,95}$Mo at $\approx 7$ MeV~(\cite{RIPL} and references therein),
and for $^{92,94,96,98,100}$Mo from $(\gamma,\gamma')$ experiments~\cite{Rusev}. The latter, however, 
shows an RSF with a shape quite different (convex rather than concave) from the one extracted from the Oslo data. 
In addition, the absolute value of the data  presented in Ref.~\cite{Rusev} appears to overestimate the experimental 
average radiative width  $\left< \Gamma_{\gamma}(S_n) \right>$. The reason for this disagreement is not yet understood, 
but the explanation might be connected
to the different reaction parameters and selectivity of the populated states 
(such as restrictions on the spin range and/or parity) compared to the Oslo data. 
Also, these data reach energies down to $E_{\gamma} \approx 5-6$ MeV only, depending on the isotope studied.
We will therefore in the following use the results from the Oslo method  to constrain the $E_\gamma \rightarrow 0$ limit.

The existence of the upbend structure could also be questioned on the basis of the various assumptions related 
to the Oslo method. In particular, one fundamental assumption behind the extraction procedure relies on the Brink 
hypothesis \cite{br55}, which states that collective excitation modes built on excited states have the same properties 
as those built on the ground state. This hypothesis allows to express the probability of the $\gamma$ decay in the 
statistical regime as being proportional to a separable product of the final-state level density and the RSF.
Although the Brink assumption can be questionable, in particular from the point of view of some models such as 
the Fermi liquid theory \cite{ka83}, it should be emphasized that the presence of the upbend structure has 
been tested against such an assumption. Specifically, the RSFs for $^{56,57}$Fe \cite{Fe_Alex}, 
$^{96,98}$Mo \cite{Mo_RSF}, and $^{45}$Sc \cite{Sc} have been determined for various initial energies 
and shown to present the upbend structure for all the excitation bins studied. From such a test, 
it can be inferred that the eventual temperature-dependence of the RSF at low energy is small with respect to the 
strength of the upbend structure. In this low-energy region, the validity of the Brink hypothesis remains 
a fundamental open question.

Uncertainties in the subtraction procedure of the Oslo method, including the estimate of the statistical multiplicity 
have also been studied in \cite{Mo_reanalyzed} and shown not to affect the RSF significantly and definitely not the 
conclusion regarding the existence of the upbend pattern.

Finally, it should also be stressed that quantitatively the experimental determination of the NLD and RSF is 
model-dependent, as the method enables a unique determination only of the \textit{functional form} of the NLD 
and RSF. In order to obtain the absolute value of the total level density (and hence the RSF) from the measured data, 
the experimental NLD needs to be normalized to the total level density at the neutron separation energy 
$S_n$, which in turn is derived from the neutron resonance spacing. The parity and spin distributions need to be 
known to estimate the total level density from the resonance spacing. As shown in Ref.~\cite{go08}, uncertainties 
within a factor of two can still affect this NLD normalization procedure and consequently could affect the low-energy RSF. 

\begin{figure*}[htb]
 \begin{center}
 \includegraphics[clip,width=1.5\columnwidth]{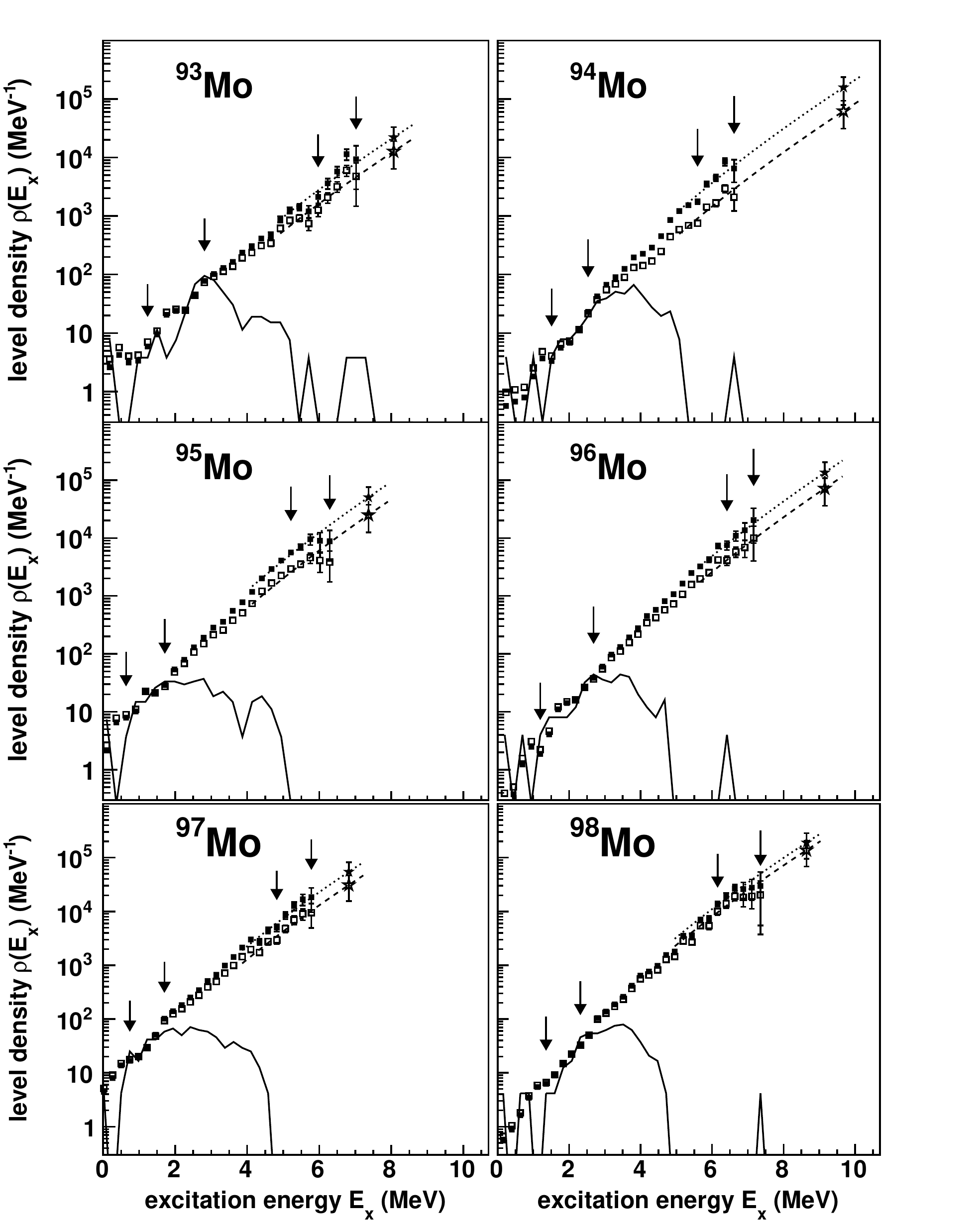} 
\vskip 2cm
 \caption{\label{fig00} Level densities for $^{93-98}$Mo with the 
	normalization of \cite{Mo_RSF} (open squares), and renormalized on the basis of the  
	level density calculations from~\cite{go08} (filled squares), 
	except for $^{98}$Mo, see text. The solid line represents the known, 
	discrete levels taken from~\cite{ENSDF}. The data points between the 
	arrows are used for normalization. The dashed line and the dotted line 
	are the interpolations between the data points and the calculated total 
	level densities at $S_n$ (open and filled star) for the original and 
	new normalization, respectively.}
 \end{center}
 \end{figure*}

\begin{figure*}[htb]
 \begin{center}
 \includegraphics[clip,width=1.5\columnwidth]{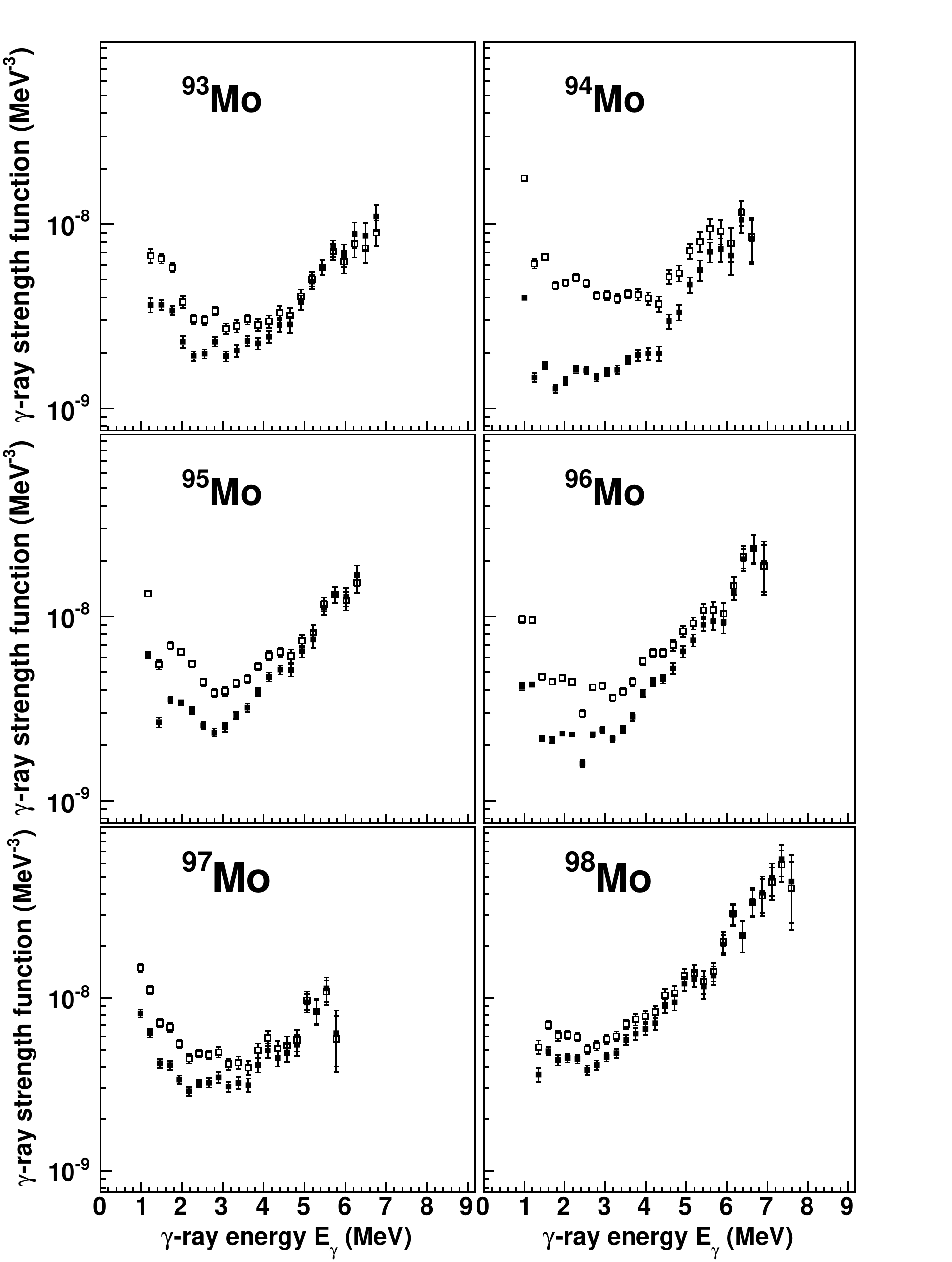}
\vskip 2cm
 \caption{\label{fig01} Gamma-ray strength functions for $^{93-98}$Mo with the 
	normalization of \cite{Mo_RSF} (open squares), and renormalized on the basis of the NLD calculations from~\cite{go08} (filled squares)
}
 \end{center}
 \end{figure*}

This uncertainty has been considered here. In Fig.~\ref{fig00}, the NLDs of $^{93-98}$Mo have been renormalized to 
the calculated total level densities at $S_n$ taken from Ref.~\cite{go08}, which  were themselves normalized to available experimental s-wave spacings. One exception is $^{98}$Mo, for which the 
newly recommended value of the s-wave spacing $D_0 = 60(10)$ eV~\cite{RIPL3} has been adopted. This new value leads to 
$\rho(S_n) = 1.38(69) \times 10^5$ MeV$^{-1}$ if calculated with the same prescription as in~\cite{Mo_RSF}, 
and to $\rho(S_n) = 1.90(95) \times 10^5$ MeV$^{-1}$ following~\cite{go08}. It is seen that both normalizations 
reproduce well the known, discrete levels taken from \cite{ENSDF}. In Fig.~\ref{fig01} the corresponding RSFs for 
both normalizations are displayed, and it is seen how the slope of the RSFs is changed. Note that for both cases 
the total RSF is normalized in absolute value to the average total radiative width $\left< \Gamma_{\gamma}(S_n) \right>$. 
For the cases of $^{93,95,97,98}$Mo the upbend structure is clearly present and relatively strong, for both
normalizations of the NLD, while for $^{94,96}$Mo the enhancement is reduced. However, this new normalization does not question the presence of the upbend in the RSF. 

In summary, the Oslo method has now been widely tested to confirm its capacity to determine the NLD and RSF. 
It has proven to be an excellent procedure and there is, at present, no reason not to trust the low-energy RSF data 
showing an upbend pattern. For that reason, we will now assume that this structure is present in all elements lighter 
than typically Cd (no upbend has been seen in Sn isotopes \cite{Sn}), and discuss how it can be modeled as an $E1$ strength.

\subsection{Model parameterizations}

To describe the $\gamma$-ray strength function with the enhancement observed at low energy, we consider different ways to model 
its contribution to the total $E1$ strength. More specifically, we have applied the widely used Generalized Lorentzian (GLO) 
model \cite{ko87,ko90} for the $E1$ strength, and modified it in order to describe the observed upbend structure. 

The GLO model makes use of a temperature 
dependence that increases the $E1$ de-excitation strength at low energies and gives an $E_\gamma \rightarrow 0$ non-zero 
limit, but no upbend pattern. This non-zero limit was first introduced by Kadmenski{\u{\i}},
Markushev and Furman in 1983 on the basis of theoretical calculations within the 
Fermi liquid model  \cite{ka83}. This was done in order to take into account quasi-particle fragmentation of the 
$E1$ strength, and to explain the observed low-energy data on the $\gamma$-ray strength function of
several medium-mass and heavy nuclei studied with $(n,\gamma\alpha)$ and $(n,\gamma f)$ reactions (see~\cite{ka83}
and references therein). Later, Kopecky and Chrien \cite{ko87}, and Kopecky and Uhl \cite{ko90} found it necessary 
to introduce a temperature dependence
in order to describe the strength of primary $\gamma$-ray data from average resonance capture (ARC)
reactions, and introduced the GLO model \cite{ko87,ko90}. This model relies also on the theory of
Fermi liquids and accounts for microscopic properties of the GEDR. 
The advantage of the GLO model is its capability to reproduce both photoabsorption cross-section data 
as well as the above-mentioned sub-threshold data reasonably well.

The GLO strength function is given by \cite{ko90}
\begin{eqnarray}
&&f_{\rm GLO}(E_{\gamma},T_f) = \frac{1}{3\pi^2\hbar^2c^2}\sigma_{E1}\Gamma_{E1} \times \\ \nonumber
&& \left[\frac{ E_{\gamma} \Gamma(E_{\gamma},T_f)}{(E_\gamma^2-E_{E1}^2)^2 + E_{\gamma}^2 
	\left[\Gamma (E_{\gamma},T_f)\right]^2} + \;0.7\frac{\Gamma(E_{\gamma}=0,T_f)}{E_{E1}^3} \right],
\label{eq:GLo}
\end{eqnarray} 
where $\sigma_{E1}$, $\Gamma_{E1}$, and $E_{E1}$ are the Giant Electric Dipole Resonance (GEDR) peak cross section, 
width, and centroid energy, respectively. The energy- and temperature-dependent width reads
\begin{equation}
\Gamma(E_{\gamma},T_f) = \frac{\Gamma_{E1}}{E_{E1}^2} (E_{\gamma}^2 + 4\pi^2T_f^2), 
\label{eq:edepwidth}
\end{equation}
identical to the prediction of \cite{ka83}. Here, the first term reflects the spreading of particle-hole states 
into more complex configurations, and the second term accounts for collisions between quasiparticles.
The spreading width thus depends on the nuclear temperature of the final states $T_f$.

\begin{figure*}[htb]
 \begin{center}
 \includegraphics[scale=0.66]{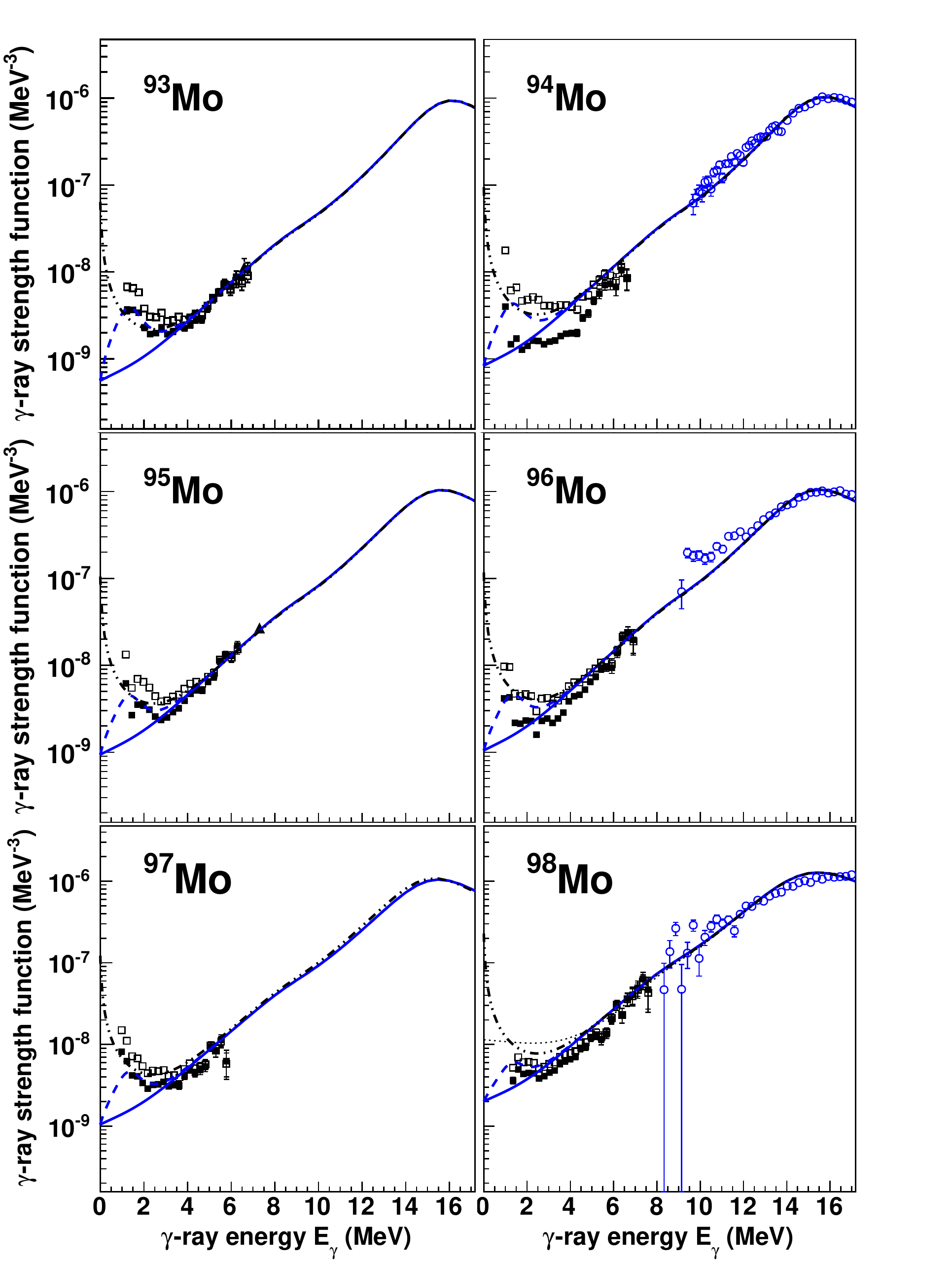}
 \caption{\label{fig02} (Color online) Gamma-ray strength functions for $^{93-98}$Mo. Experimental data points 
	with the normalization of Ref.~\cite{Mo_RSF} are shown as open squares. The filled squares are obtained 
	when normalizing the experimental NLDs on the basis of the calculations of~\cite{go08}. Giant resonance photoabsorption 
	data (blue open circles) for $^{94,96,98}$Mo are taken from \cite{Beil}. The black triangles represent 
	measured $E1$ strengths for $^{93,95}$Mo from~\cite{RIPL}. The blue solid line corresponds to the GLO-lo 
	parameterization, the blue dashed line to the GLO-up1 parameterization, and the dash-dot line shows the GLO-up2 model.
	For $^{98}$Mo, also the GLO model for $E_n=1$ MeV is displayed (dotted line).}
 \end{center}
 \end{figure*}

The Oslo method is based on the Brink hypothesis, which means no dependence on the excitation energy and thus on the 
nuclear temperature $T_f$ of final states in the RSF. Introducing a constant temperature is, however, not in contradiction 
with the Brink hypothesis or the extraction procedure to get the NLD and RSF from the coincidence data. 
We have therefore considered $T_f$ to be constant in the GLO model. We have chosen $T_f = 0.30$ MeV, 
to  give a reasonable global 
agreement with experimental data of $^{93-98}$Mo for $E_{\gamma} \gtrsim 3$ MeV, as illustrated in Fig.~\ref{fig02}; 
this first adaptation of the GLO model is in the following referred to as GLO-lo.   

For the GEDR parameters we take experimental values from \cite{RIPL,iaea00} 
when available (for $^{98}$Mo a new improved determination has been performed), and interpolated values from the 
even-even $^{92,94,96,98}$Mo  for the missing odd isotopes $^{93,95,97}$Mo. For the heavier Mo nuclei ($A>98$), 
we used the systematics recommended in Ref.~\cite{RIPL}.  The final GEDR parameters for $^{93-98}$Mo are summarized  in Table~\ref{tab:parameters}.
\begin{table}[htb]
\caption{Parameters used for the GEDR strength of $^{93-98}$Mo.} 
\begin{tabular}{lccc}
\hline
\hline
Nucleus    & $E_{E1}$ & $\sigma_{E1}$ & $\Gamma_{E1}$     \\
           & (MeV)  		 & (mb)  &  (MeV) \\
\hline
$^{93}$Mo  & 16.59 & 173.5 & 4.82 \\
$^{94}$Mo  & 16.36 & $185.0$ & 5.50 \\
$^{95}$Mo  & 16.28 & $185.0$ & 5.76 \\
$^{96}$Mo  & 16.20 & $185.0$ & 6.01 \\
$^{97}$Mo  & 16.00 & $187.0$ & 5.98 \\
$^{98}$Mo  & 16.50 & $ 220.0$ & 8.00 \\
\hline
\hline
\end{tabular}
\\
\label{tab:parameters}
\end{table}

We have chosen two ways to model the upbend:
\begin{itemize}
\item[(\textit{i})]{introducing a low-lying resonance of the form of a standard Lorentzian (SLO).}
\item[(\textit{ii})]{modifying the energy-dependent width of the GLO model.}
\end{itemize}
It should be stressed that both approaches are completely phenomenological, since there is at present 
no proper theoretical description of the upbend structure. 

For the first approach, the upbend is given by a low-lying resonance described by an SLO shape:
\begin{equation}
f_{\mathrm{up1}}(E_\gamma)=\frac{1}{3\pi^2\hbar^2c^2} 
	\frac{\sigma_{\mathrm{up1}}E_\gamma\Gamma_{\mathrm{up1}}^2} {(E_\gamma^2-E_{\mathrm{up1}}^2)^2+E_\gamma^2\Gamma_{\mathrm{up1}}^2},
\label{eq:lorup}
\end{equation}
located at a resonance energy of $E_{\mathrm{up1}}=1.5$~MeV with a full width at half maximum 
$\Gamma_{\mathrm{up1}}=1.5$~MeV and a peak cross section $\sigma_{\mathrm{up1}}=0.05$~mb. This resonance  
is added to the GLO-lo strength in order to make the total
RSF fit with the low-energy data as well; this model is referred to as GLO-up1 and its Mo description is shown in Fig.~\ref{fig02}.

For the second approach, we have modified the temperature-dependent width of the GLO model in the 
following way:
\begin{equation}
\Gamma_{\rm up2}(E_{\gamma},T_f) = \frac{\Gamma_{E1}}{E_{E1}^2} \left[E_{\gamma}^2 + 
	\frac{4\pi^2T_f^2E_{E1}}{(E_{\gamma}+\delta)}\right], 
\label{eq:glowidthmod}
\end{equation}
where the introduction of an $E_{\gamma}^{-1}$ dependence in the second term enables an increasing
RSF for decreasing $E_{\gamma}$. The constant parameter $\delta = 0.05$ MeV is applied to ensure a finite value
of $f_{\rm GLO}$ for $E_\gamma \rightarrow 0$. In this approach, the temperature $T_f $ has also been assumed to remain constant and the value of $T_f= 0.16$ MeV has been adjusted to reproduce at best the experimental Mo RSF, as seen in Fig.~\ref{fig02}.  Note that the $T_f$ value is obviously different from the value extracted wihtin the GLO-lo model because of the new functional (\ref{eq:glowidthmod}) considered for the energy-dependent width.
This modified width  allows us to use for the RSF a closed form identical to Eq.~(\ref{eq:GLo}) that makes it possible to describe the RSF at all energies and also to reproduce the upbend structure 
phenomenologically without calling for the presence of an extra low-lying resonance. We call this model GLO-up2.

Since most of the reaction calculations are performed with the original version of the GLO model \cite{ko90}, 
we also consider the corresponding model with a variable temperature $T_f$, estimated from the well-known expression \cite{RIPL}
\begin{equation}
T_f = \sqrt{(E^* -\Delta - E_\gamma)/a},
\label{eq:temp}
\end{equation}
where $E^*$ is the initial excited state in the compound nucleus (for neutron capture at incoming neutron energy 
$E_n$, $E^*=E_n+S_n$), $\Delta$ is a 
pairing correction, and $a$ is the level density parameter at $S_n$.

For all $E1$ strength functions considered here, we have also added an $M1$ strength described by a Lorentzian shape:
\begin{equation}
f_{M1}(E_\gamma)=\frac{1}{3\pi^2\hbar^2c^2} 
	\frac{\sigma_{M1}E_\gamma\Gamma_{M1}^2} {(E_\gamma^2-E_{M1}^2)^2+E_\gamma^2\Gamma_{M1}^2}.
\label{eq:M1}
\end{equation}
Here,  $\sigma_{M1}$, $\Gamma_{M1}$, and $E_{M1}$ are the peak cross section, 
width, and centroid energy, respectively, of the Giant Magnetic Dipole Resonance (GMDR) related to spin-flip transitions
between major shells~\cite{bohr&mottelson}. The peak cross section is normalized to the $E1$ strength 
function at $E_{\gamma} = 7$ MeV as described in Ref.~\cite{RIPL}, while the width and the peak position
is determined from systematics~\cite{RIPL}. With such a parameterization the $M1$ contribution to the total RSF remains low compared to the $E1$ contribution.

The GLO-lo, GLO-up1 and GLO-up2 models are 
shown together with experimental data on $^{93-98}$Mo in Fig.~\ref{fig02}. For $^{98}$Mo, also the GLO model
with $E_n = 1$ MeV is displayed. It is seen that both the GLO-up1 and the GLO-up2 models give reasonable agreement with  data at all energies, while GLO-lo fails to describe the low-energy region. 

In the following section, the impact of the upbend structure is estimated by comparing the above-mentioned models (GLO-up1,
GLO-up2, and GLO-lo) and the original and widely used $T$-dependent GLO model.

\section{Results}
\label{sect_res}

We now perform the calculation of $(n,\gamma)$ cross-sections and astrophysical rates with  the code 
TALYS \cite{TALYS,go08b} in order to study the impact the upbend structure might have on the radiative 
neutron capture for stable as well as neutron-rich nuclei. By default, all calculations are performed 
using the nuclear structure properties determined within the HFB-17 mass model \cite{go09}, the 
neutron-nucleus optical potential of Ref.~\cite{ko03} and the NLD obtained within the combinatorial 
method \cite{go08}. The latter model not only reproduces quite accurately resonance spacing data, but 
also the energy-dependence of the total level density extracted consistently for the Mo isotopes through 
the Oslo method. The Oslo data and the theoretical NLD are compared in Fig. 9 of Ref.~\cite{go08}. 
To remain coherent, it is of prime importance to use the same NLD prescription and normalizing values at 
the neutron separation energy for the cross section calculation as those used in the extraction of the RSF 
with the Oslo method. As shown in Fig.~\ref{fig01}, different NLD models lead to different experimental RSFs. 

 \begin{figure}[htb]
 \begin{center}
 \includegraphics[clip,scale=0.32]{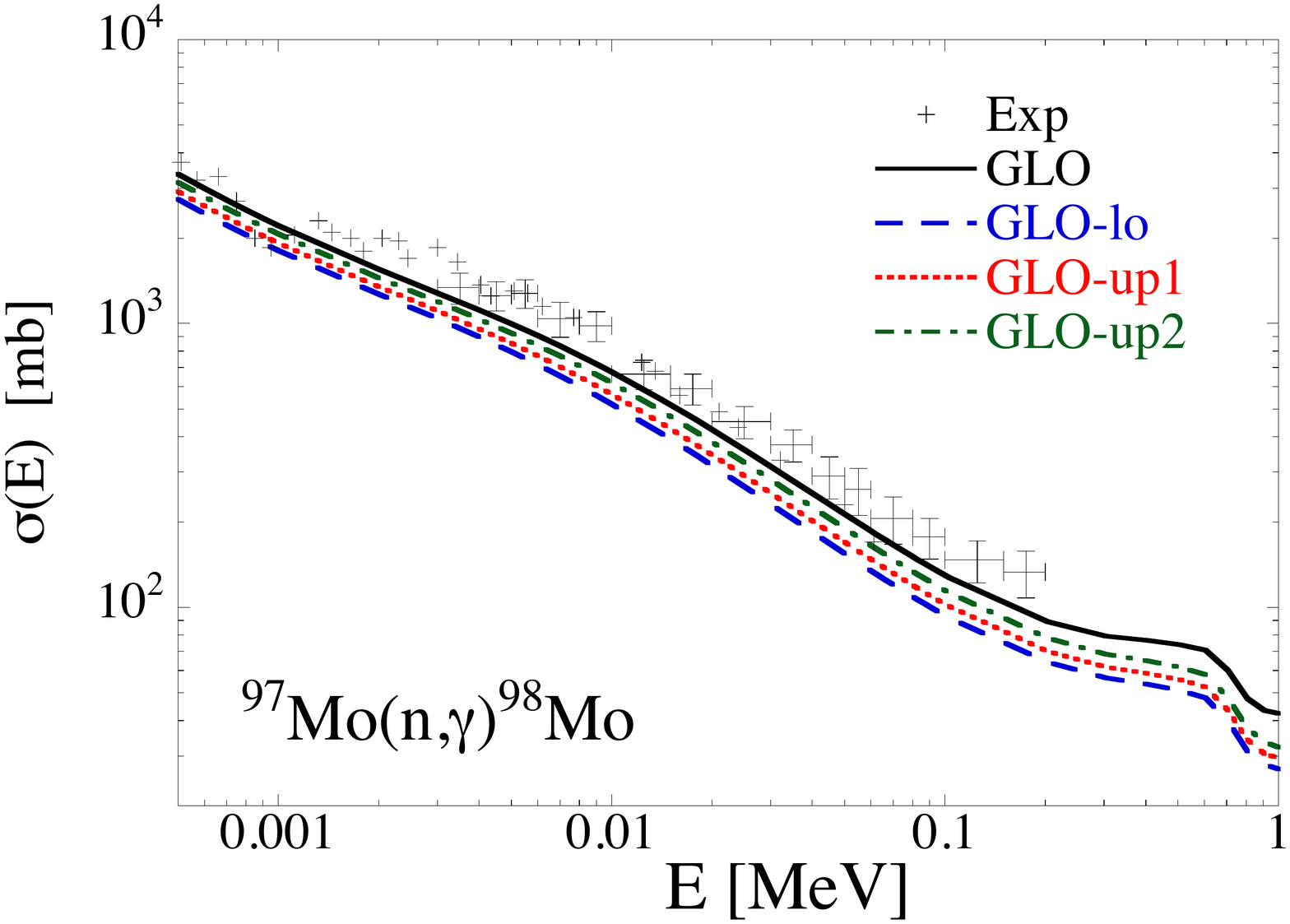}
  \caption{\label{fig04} (Color online) Comparison of experimental \cite{ka64,mu76} and calculated 
	$^{97}$Mo$(n,\gamma)^{98}$Mo cross section obtained with the GLO (solid line), GLO-lo (blue dashed line), 
	GLO-up1 (red dotted line) and GLO-up2 (green, dash-dot line) models. }
 \end{center}
 \end{figure}

We compare in Fig.~\ref{fig04} experimental $^{97}$Mo$(n,\gamma)^{98}$Mo data with the calculated cross section 
obtained with the original GLO model, the GLO-lo model with $T_f=0.3$ MeV, and the two parameterizations of the upbend, GLO-up1 and
GLO-up2. It is seen from Fig.~\ref{fig04} that the increase in the predicted cross section using our description 
of the upbend structure for this specific stable Mo isotope can reach about 20\% if we adopt the GLO-up1 
parameterization with respect to the GLO-lo one, and roughly 50\% if we adopt the GLO-up2 model.  
Similar results are obtained for the other Mo isotopes. In general, the upbend structure 
improves the agreement with experimental data, although there is clearly some strength missing, partially due to 
a possible extra contribution lying in the 8-10~MeV region of $^{98}$Mo that is not properly described by the present 
parameterizations (see Fig.~\ref{fig02}). Note, however, that no effort is done here to reproduce 
experimental data perfectly, for example by modifying the NLD model or the RSF parameterization. 
When considering 
the original GLO model,
the estimated cross section is slightly higher due to the high temperatures $T_f$ encountered in the compound system in comparison 
with the constant values adopted.

\begin{figure}[htb]
 \begin{center}
 \includegraphics[clip,scale=0.32]{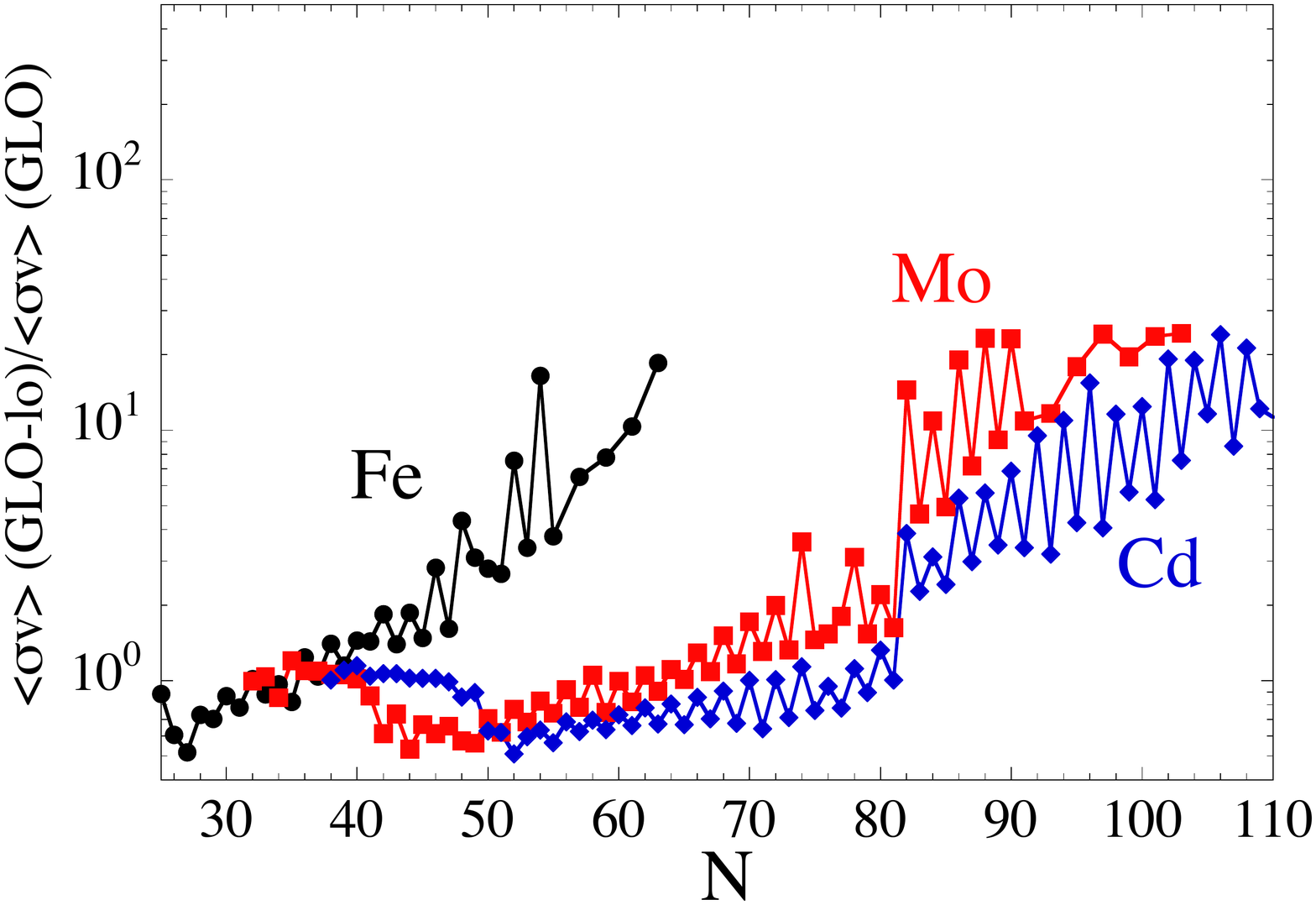}
  \caption{\label{fig05} (Color online) Ratios of Maxwellian-averaged $(n,\gamma)$ reaction rates at 
	$T=10^9$~K for the Fe, Mo and Cd isotopic 
	chains up to the neutron drip line, using the GLO-lo and GLO model.}
 \end{center}
 \end{figure}

\begin{figure}[htb]
 \begin{center}
 \includegraphics[clip,scale=0.32]{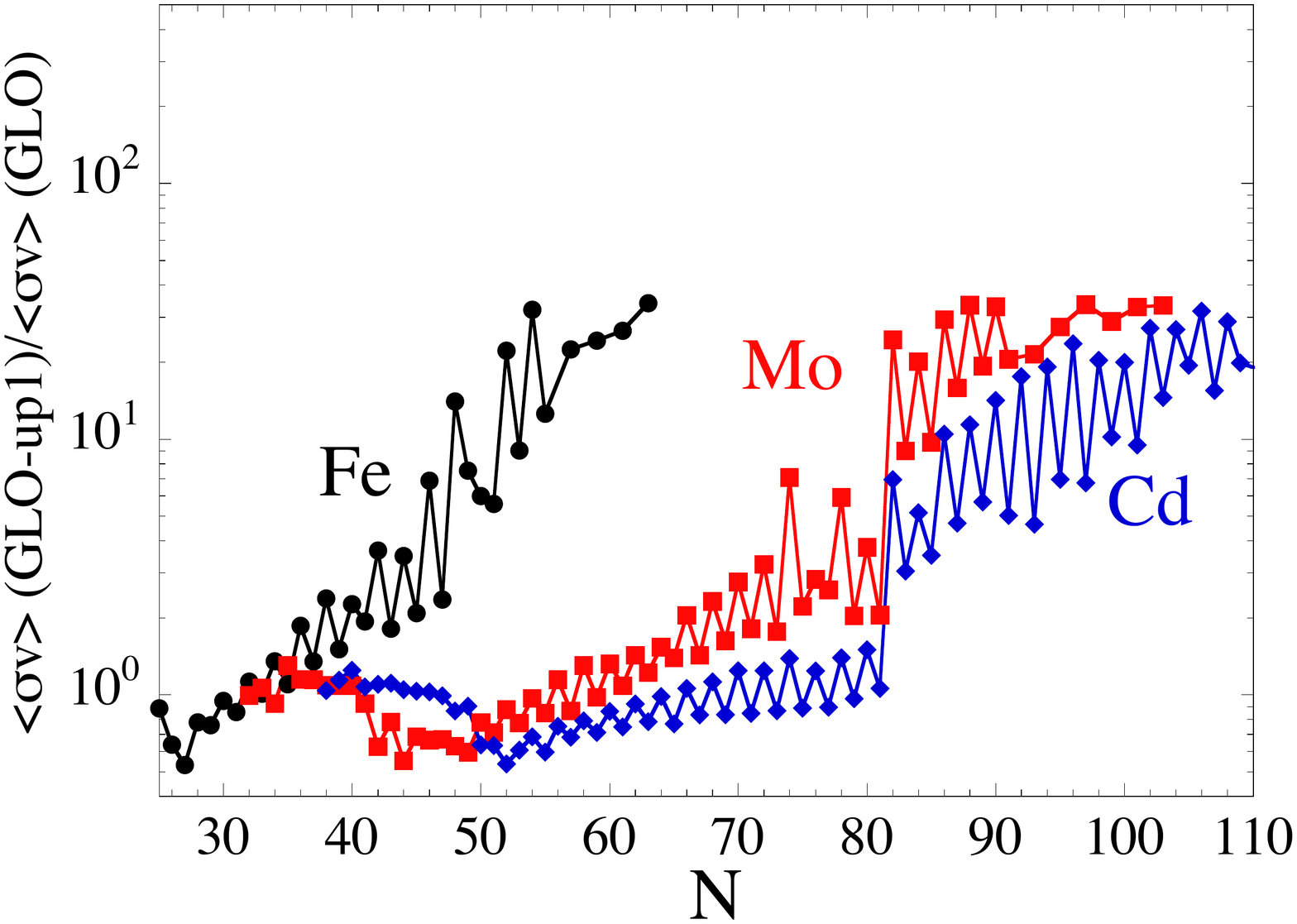}  
  \caption{\label{fig06} (Color online) Same as Fig.~\ref{fig05} for the GLO-up1 and the GLO model.}
 \end{center}
 \end{figure}

\begin{figure}[htb]
 \begin{center}
 \includegraphics[clip,scale=0.32]{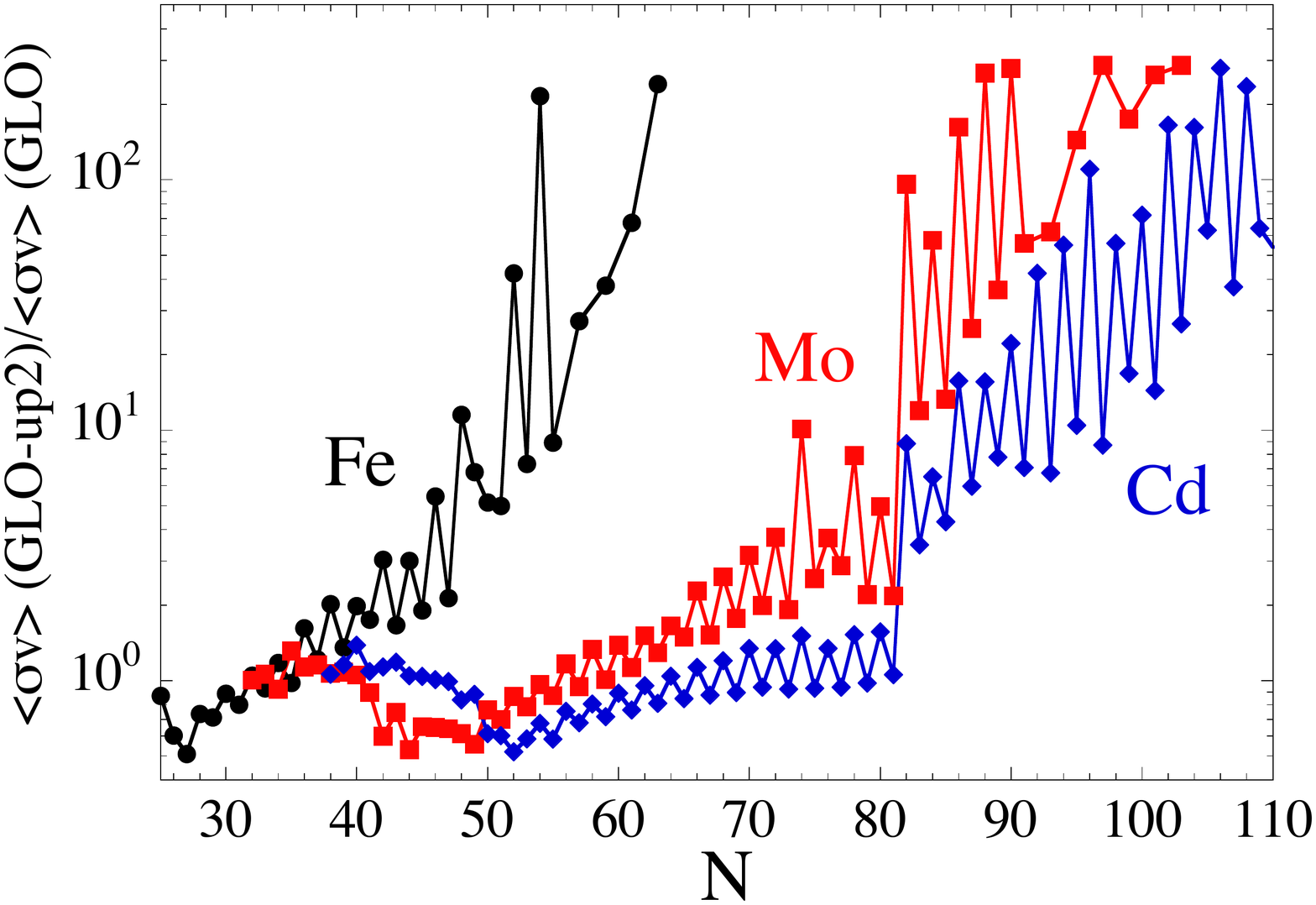}  
  \caption{\label{fig07} (Color online) Same as Fig.~\ref{fig05} for the GLO-up2 and the GLO model.}
 \end{center}
 \end{figure}

On the basis of the input models described above, we now perform calculations on the Maxwellian-averaged neutron capture rates of 
astrophysical interest for the full isotopic chains of Mo, as well as Fe and Cd up to the neutron drip line. 
The GLO-up1, GLO-up2, GLO-lo predictions are compared with the widely used GLO estimates in Figs.~\ref{fig05} -- \ref{fig07} for a temperature of $T=10^9$~K typical of the r-process nucleosynthesis \cite{ar07}. As already demonstrated in Fig.~\ref{fig04}, 
close to the stability line the upbend structure has a relatively small influence. However, for exotic 
neutron-rich nuclei the impact may become large, essentially due to the low neutron separation energies allowing 
only for $\gamma$-decays with energies lower than typically 2~MeV. In this case, the strength in the low-energy region 
dominates the decay. 

In Fig.~\ref{fig05}, the rates obtained using the GLO-lo with constant temperature of $T_f=0.3$ MeV are compared to 
the frequently used standard GLO model with variable temperature as defined in Eq.~(\ref{eq:temp}). For nuclei close to 
the valley of stability, it is seen that 
the constant-temperature approach gives lower rates than the original GLO model, which is easy to understand from the
higher absolute value of the GEDR tail when using a variable temperature which is found to be higher than the 0.3~MeV 
considered in the GLO-lo model (see Fig.~\ref{fig02} for $^{98}$Mo). However, 
when approaching the neutron drip line, the rates of the GLO-lo model become comparable and even larger than the ones
using the GLO model. This is due to the fact that the neutron separation energy drops rapidly and so does the temperature $T_f$, at least for neutron incident energies of about 100~keV (corresponding to the $T=10^9$~K temperature considered here). The original GLO model when applied to the neutron capture by exotic neutron-rich nuclei can therefore be approximated by the $T_f=0$ case. We see from Fig.~\ref{fig05} that assuming a constant temperature $T_f=0.3$~MeV (i.e the GLO-lo case) can give an order of magnitude  increase in the reaction rates for such exotic nuclei. 

Including the upbend structure through the GLO-up1 model may give another significant increase of the rates as shown 
in Fig.~\ref{fig06}. In particular, the rates for neutron-rich Cd isotopes gain an additional order of magnitude due 
to the low-energy RSF contribution that become effective as soon as $S_n$ drops after crossing the 
closed neutron shell at $N=82$. As demonstrated in Fig.~\ref{fig07}, the GLO-up2 parameterization 
gives a similar large increase of the rates with respect to the traditional calculation based on the GLO model. 
The predictions are even larger than when considering the GLO-up1 model. Similar conclusions can be drawn for the Fe and 
Mo isotopes.

In general, we see that the influence of the upbend structure on the $(n,\gamma)$ cross sections and thus the reaction rates 
becomes more and more 
important as the number of neutrons increases. In particular, as soon as a major neutron shell is crossed, the neutron 
separation energy $S_n$ drops and the RSF in the vicinity of the upbend structure starts to play a major role in 
the radiative decay. It can be seen that the combination of the upbend structure and applying a constant temperature 
may lead to an increase of the reaction rates by up to a 
factor of 300. This increase is observed in all the isotopic chains studied here when applying the GLO-up2 model.

These calculations show that a proper understanding of the $E_\gamma \rightarrow 0$ limit of the RSF can be of 
crucial importance in the determination of radiative neutron-capture cross sections for exotic neutron-rich nuclei. 
This effect could have a non-negligible impact on the neutron captures that can potentially take place in 
astrophysical environments characterized by high neutron densities, in particular during the r-process nucleosynthesis.

\section{Conclusions}
\label{sect_conc}

As shown experimentally by the Oslo group, the RSF at very low $\gamma$-ray energy might be characterized by 
a significant enhancement with respect to the usual rapidly decreasing $\gamma$-decay strength. This  
upbend structure has been observed systematically in V, Sc, Fe and Mo isotopes at energies typically smaller 
than 3~MeV, but is absent in elements heavier than Sn. The very origin of this extra strength remains unexplained 
theoretically. 

The impact of this upbend structure is found to be relatively small on the neutron capture cross 
section of stable nuclei, since it modifies the RSF in an energy region which hardly takes part in the reaction 
mechanism. However, for exotic neutron-rich nuclei, this effect becomes significant and could 
potentially increase the reaction rates of astrophysical relevance by one or even two orders of magnitude. 
This effect is particularly pronounced for nuclei with a low neutron separation energy, crossing a major 
neutron shell. This effect could have a non-negligible impact on the neutron capture rates essential for the 
r-process nucleosynthesis.

\vfill

\begin{thebibliography}{99}
\bibitem{ar07} M. Arnould, S. Goriely, and K. Takahashi,  Phys. Rep.  {\bf 450}, 97 (2007).
\bibitem{go98} S. Goriely,  Phys. Lett. {\bf B436}, 10 (1998).
\bibitem{go96} S. Goriely, M. Arnould, Astron. Astrophys. {\bf 312} 327 (1996). 
\bibitem{Fe_Alex} A.~Voinov, E.~Algin, U.~Agvaanluvsan, T.~Belgya, R.~Chankova,
	M.~Guttormsen, G.E.~Mitchell, J.~Rekstad, A.~Schiller and S.~Siem, Phys. Rev. Lett {\bf 93}, 142504 (2004).
\bibitem{Fe_Emel} E.~Algin, U.~Agvaanluvsan, M.~Guttormsen, A.~C.~Larsen, G.~E.~Mitchell, 
	J.~Rekstad, A.~Schiller, S.~Siem, and A.~Voinov, Phys. Rev. C {\bf 78}, 054321 (2008).
\bibitem{Mo_RSF} M.~Guttormsen, R.~Chankova, U.~Agvaanluvsan,  E.~Algin, L.A.~Bernstein, 
	F.~Ingebretsen, T.~L{\"o}nnroth, S.~Messelt, G.E.~Mitchell, J.~Rekstad, A.~Schiller, 
	S.~Siem, A.C.~Sunde, A.~Voinov and S.~{\O}deg{\aa}rd,
	Phys. Rev. C {\bf 71}, 044307 (2005).
\bibitem{V} A.~C.~Larsen, R.~Chankova, M.~Guttormsen, F.~Ingebretsen, T.~L{\"o}nnroth, 
	S.~Messelt, J.~Rekstad, A.~Schiller, S.~Siem, N.~U.~H.~Syed, A.~Voinov, and 
	S.~W.~{\O}deg{\aa}rd, Phys.\ Rev.\ C \bf 73\rm, 064301 (2006).
\bibitem{Sc} A.~C.~Larsen, M.~Guttormsen, R.~Chankova, F.~Ingebretsen, T.~L{\"o}nnroth, 
	S.~Messelt, J.~Rekstad, A.~Schiller, S.~Siem, N.~U.~H.~Syed, and A.~Voinov, 
	Phys.\ Rev.\ C \bf 76\rm, 044303 (2007).
\bibitem{PragueM1} M.~Krti\u{c}ka, F.~Be\u{c}v\'{a}\u{r}, J.~Honz\'{a}tko, I.~Tomandl, 
	M.~Heil, F.~K\"{a}ppeler, R.~Reifarth, F.~Voss, K.~Wisshak, Phys.~Rev.~Lett.~{\bf 92}, 172501 (2004).
\bibitem{SchillerM1} A.~Schiller, A.~Voinov, E.~Algin, J.~A.~Becker, L.~A.~Bernstein, P.~E.~Garrett, 
	M.~Guttormsen, R.~O.~Nelson, J.~Rekstad, S.~Siem, Phys.~Lett.~B {\bf 633}, 225 (2006).
\bibitem{Schiller00} A.~Schiller, L. Bergholt, M.~Guttormsen, E. Melby, J.~Rekstad, S.~Siem, 
	Instrum. Methods Phys. Res. A {\bf 447} 494 (2000).
\bibitem{Fransen1} C.~Fransen {\textit{et al}}., Phys. Rev. C \bf 67\rm, 024307 (2003). 
\bibitem{Fransen2} C.~Fransen {\textit{et al}}., Phys. Rev. C \bf 70\rm, 044317 (2004).
\bibitem{Alex_Ni}  A.~Voinov, S.~M.~Grimes, C.~R.~Brune, M.~Guttormsen, A.~C.~Larsen, 
	T.~N.~Massey, A.~Schiller, and S.~Siem, Phys. Rev. C \bf 81\rm, 024319 (2010).
\bibitem{RIPL} T.~Belgya, O.~Bersillon, R.~Capote, T.~Fukahori, G.~Zhigang, S.~Goriely, 
	M.~Herman, A.~V.~Ignatyuk, S.~Kailas, A.~Koning, P.~Oblozinsky, V.~Plujko and P.~Young, 
	\textit{Handbook for calculations of nuclear reaction data}, RIPL-2. \textbf{IAEA-TECDOC-1506} 
	(IAEA, Vienna, 2006); also  available online at http://www-nds.iaea.org/RIPL-2/ 
\bibitem{Rusev} G.~Rusev et al., Phys.~Rev.~C {\bf 79} 061302 (2009).
\bibitem{br55} D.~M. Brink, PhD thesis, Oxford University (1955).
\bibitem{ka83} S.~G.~Kadmenski{\u{\i}}, V.P. Markushev, and V.I. Furman, Sov. J. Nucl. Phys.  {\bf 37}, 165 (1983).
\bibitem{Mo_reanalyzed}Ê M.~Guttormsen, R.~Chankova, U.~Agvaanluvsan,  E.~Algin, 
	L.~A.~Bernstein, F.~Ingebretsen, T.~L{\"o}nnroth, S.~Messelt, G.E.~Mitchell, 
	J.~Rekstad, A.~Schiller, S.~Siem, A.C.~Larsen, A.~Voinov and S.~{\O}deg{\aa}rd, Los Alamos preprint server: \\ 
	\url{http://xxx.lanl.gov/abs/0801.4667}.
\bibitem{go08} S. Goriely, S. Hilaire, and A.J. Koning, Phys. Rev. C \textbf{78}, 064307 (2008).
\bibitem{RIPL3} R. Capote, et al., 
Nuclear Data Sheets, {\bf 110}, 3107 (2009). See also \emph{http://www-nds.iaea.org/RIPL-3/.}

\bibitem{ENSDF}Data extracted using the NNDC On-Line Data Service from the ENSDF database.
\bibitem{Sn} U.~Agvaanluvsan, A.~C.~Larsen, R.~Chankova, M.~Guttormsen, G.~E.~Mitchell, 
	A.~Schiller, S.~Siem, and A.~Voinov, Phys.\ Rev.\ Lett. \bf 102 \rm, 162504 (2009).
\bibitem{ko87} J. Kopecky and R.~E.~Chrien, Nucl. Phys. \bf A468\rm, 285 (1987).
\bibitem{ko90} J. Kopecky and M. Uhl, Phys.~Rev.~C {\bf 41}, 1941 (1990).
\bibitem{iaea00} Photonuclear data for applications; cross sections and spectra 2000,
IAEA-Tecdoc-1178.
\bibitem{bohr&mottelson} A.~Bohr and B.~Mottelson, {\em Nuclear Structure}, (Benjamin,
New York, 1975), Vol.~II, p.~636.
\bibitem{TALYS} A.~J.~Koning, S.~Hilaire, and M.~C.~Duijvestijn, 
	\textit{"TALYS-1.0"}, in Proceedings of the International Conference on Nuclear Data 
	for Science and Technology, April 22-27, 2007, Nice, France. 
	Editors: O.~Bersillon, F.~Gunsing, E.~Bauge, R.~Jacqmin, and S.~Leray, EDP Sciences, 211 (2008). 
\bibitem{go08b} S. Goriely,  S. Hilaire, A.J. Koning, \emph{Astron. Astrophys.}, \textbf{487}, 767 (2008).
\bibitem{go09} S. Goriely, N. Chamel, and J.M. Pearson, Phys. Rev. Lett. \textbf{102}, 152503 (2009).
\bibitem{ko03} A.~J. Koning and J.-P. Delaroche, Nucl. Phys. {\bf A713},  231 (2003).
\bibitem{Beil} H.~Beil,~R. Berg\`{e}re, P.~Carlos, A.~Lepr\^{e}tre, A.~De~Miniac, 
	A.~Veyssi\`{e}re, Nucl. Phys. \textbf{A227}, 427 (1974).
\bibitem{ka64} S.¬P. Kapchigashev, Yu.~P. Popov, Neutron Interactions Conf., Dubna, p.104 (1964).
\bibitem{mu76} A.~R.~Del.~Musgrove, B.~J. Allen, J.~W. Boldeman, R.~L. Macklin, Nucl. Phys. {\bf A270}, 108 (1976).



\end{thebibliography}
\end{document}